\documentclass[printer]{aa}

\bibpunct{(}{)}{;}{a}{}{,}

\usepackage{graphicx}
\usepackage{txfonts}
\usepackage{soul}
\usepackage{txfonts}
\usepackage{color}
\usepackage{ulem}

\newcommand{\hii}{H\textsc{ii}}
\def\ks{km s$^{-1}$}

\def\s{$^{\prime\prime}$}

\def\cm3{cm$^{-3}$}

\def\2{$^{12}$CO}
\def\3{$^{13}$CO}
\def\8{C$^{18}$O}
\def\msol{M$_\odot$}

\def\cm2{cm$^{-2}$}

\begin{document}

\title{Multiple molecular outflows and fragmentation in the IRDC core G34.43+00.24 MM1}

\author {N. L. Isequilla \inst{1}
\and M. E. Ortega \inst{1}
\and M. B. Areal \inst{1}
\and S. Paron \inst{1}
}

\institute{CONICET - Universidad de Buenos Aires. Instituto de Astronom\'{\i}a y F\'{\i}sica del Espacio
             CC 67, Suc. 28, 1428 Buenos Aires, Argentina\\
             \email{nisequilla@iafe.uba.ar}
}

\offprints{N.L. Isequilla}

   \date{Received <date>; Accepted <date>}

\abstract{ The fragmentation of a molecular cloud that leads to the formation of high-mass stars occurs on a hierarchy of different spatial scales. The large molecular clouds harbour massive molecular clumps with massive cores embedded in them. The fragmentation of these cores may determine the initial mass function and the masses of the final stars. Therefore, studying the  fragmentation processes in the cores is crucial to understand how massive stars form.}
{ Detailed studies towards particular objects are needed to collect observational evidence that shed light on the star forming processes at the smallest spatial scales.  
The hot molecular core G34--MM1, embedded in the filamentary infrared dark cloud (IRDC) G34.34$+$00.24 located at a distance of 3.6 kpc, is a promising object to study both the fragmentation and outflow processes.}    
{Using data at 93 and 334 GHz obtained from the Atacama Large Millimeter Array (ALMA) database we studied with great detail the hot molecular core G34--MM1. The angular resolution of the data at 334 GHz is about 0\farcs8, which allow us to resolve structures of about 0.014 pc ($\sim$2900 au).}
{We found evidence of fragmentation towards the molecular hot core G34--MM1 at two different spatial scales. The dust condensation MM1--A (about 0.06~pc in size) harbours three molecular subcores candidates (SC1 through SC3) detected in $^{12}$CO J=3--2 emission, with typical sizes of about 0.02~pc and an average spatial separation among them of about 0.03~pc. From the HCO$^+$ J=1--0 emission, we identify, with better angular resolution than previous observations, two perpendicular molecular outflows arising from MM1--A.  We suggest that subcores SC1 and SC2, embedded in MM1--A, harbour the sources responsible of the main and the secondary molecular outflow, respectively. Finally, from the radio continuum emission at 334~GHz, we marginally detected another dust condensation, named MM1--E, from which a young (${\rm t_{dyn}} \sim 1.6 \times 10^3$~yrs), massive (M $\sim$ 5 M$_{\odot}$) and energetic  (E $\sim 6 \times 10^{46}$~ergs)  molecular outflow arises.} 
{The fragmentation of the hot molecular core G34--MM1 at two different spatial scales, together with the presence of multiple molecular outflows associated with it, would support a competitive accretion scenario. Studies like this, point to shed light on the relation between fragmentation and star formation processes occurring within hot molecular cores which are only accessible through high-angular resolution interferometric observations.}

\titlerunning{Outflows and fragmentation in the core G34-MM1}
\authorrunning{N.L. Isequilla et al.}


\maketitle
%

\section{Introduction}

Given that massive stars  (M > 8~\msol) form deeply in the molecular clouds and that they reach the pre-main sequence
stage quite before the dense envelope is removed, the formation and presence of young massive stars are usually inferred from the effects that they have on the surrounding interstellar material. The presence of hot molecular cores (HMCs), ultra-compact (UC) \hii~regions, and massive molecular outflows indicate that high-mass star formation is indeed taking place \citep{rathborne11}.  

 HMCs  are compact ($\leq$ 0.1~pc), dense ($\sim 10^5 - 10^8$ cm$^{-3}$), and massive ($\sim 10^2$~\msol) molecular structures that are heated (>100 K) in the immediate vicinity of the recently formed high-mass protostar \citep{cesa05,beu07,motte18}. Additionally, HMCs are the most important
reservoirs of complex organic molecules (COMs), including key species for prebiotic
processes \citep{beltran18a}, and thus, they are very important interstellar objects in the contribution of our
knowledge about astrochemistry \citep{jor20}.  While some HMCs are associated with UC \hii~regions, many have weak or no observable radio continuum emission \citep{hoare07}. Hence, HMCs are thought to be the precursors of UC \hii~and thus, appropriate features to study the earliest 
phases of high-mass star formation and the related chemistry. Molecular outflows have also been observed during the HMC phase, which suggests the presence of accretion disks within the cores \citep{her14}. 

HMCs are usually embedded in  massive molecular clumps, that in turn are substructures of infrared dark clouds (IRDCs), which are identified as high-extinction filamentary features against the bright Galactic mid-IR background \citep{rathborne06,ragan15,lu2018}.
 This hierarchical fragmentation seems to be an ubiquitous process in the massive star formation that occurs from the scale of the filaments down to the scale of individual dense cores and protostars \citep{sada20}. 
It is thought that cores tend to further fragment, indicating that massive stars form in close groups \citep{liu13,zhang15}. Thus, studying the deepest level of fragmentation is crucial to understand not only the formation of individual stars but also of stellar clusters. 

In this work, based on high-resolution data extracted from the ALMA Science Archive, we present a kinematic and morphological study of the hot molecular core G34--MM1 in order to characterize in detail its internal structure. The paper is organized as follows:
Sect.\,\ref{present} presents the IRDC and the analyzed hot molecular core, Sect.\,\ref{dataS} describes the used data, in Sects.\,\ref{results} and \,\ref{discuss} we present and discuss the results, and finally, Sect.\,\ref{concl} states our concluding remarks.

 \begin{figure}[h]
   \centering
   \includegraphics[width=9cm]{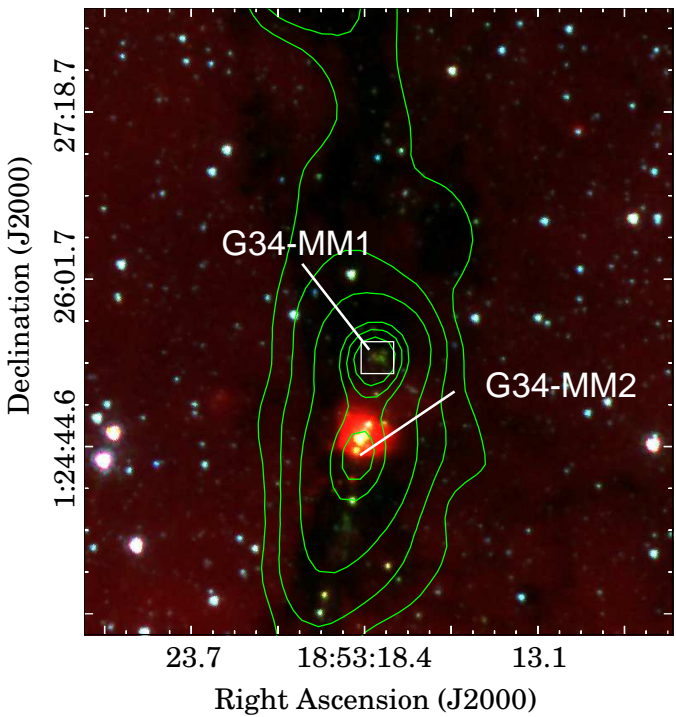}
    \caption{Overview of the IRDC G34.43$+$00.24 region containing cores G34--MM1 and G34--MM2 on a GLIMPSE three-colour image with 3.6 (blue), 4.5 (green) and 8.0~$\mu$m (red). The green contours represent the ATLASGAL emission at 870~$\mu$m. Levels are at 0.4, 0.7, 1.5, 4, 5, and 6~Jy beam$^{-1}$. The white square indicates the region mapped at 334~GHz.}
              \label{spitzer}
    \end{figure}

\section{Presentation of the region and the analyzed core}
\label{present}

IRDC G34.43+00.24 (hereafter IRDC G34) is a filamentary infrared dark cloud located at a kinematic distance of about 3.6~kpc \citep{tang19} harboring several molecular cores, named from G34-MM1 to G34-MM9, according to \citep{rathborne06}, that are likely at different evolutionary stages \citep{chen11}. For instance, among these cores, G34-MM2 has an associated UC\hii~region \citep{shepherd04}. The authors also detected marginal radio continuum emission towards G34-MM1, and suggested that the embedded object appears to be a massive B2 protostar at an early stage of evolution, which is supported by the absence of associated near-infrared emission. Later, \citet{rosero16} confirmed radio continuum emission at 1.3 and 6~cm towards G34-MM1.  

Figure \ref{spitzer} shows a three-colour {\it Spitzer} image of a region of the IRDC G34 (3.6~$\mu$m=blue, 4.5~$\mu$m=green, and 8.0~$\mu$m=red), seen as a dark filament against the 8~$\mu$m diffuse emission. The green contours represent the continuum emission at 870~$\mu$m obtained from ATLASGAL. It can be noticed two striking dust condensations, ATLASGAL sources G034.4112+0.2344 and G034.4005+0.2262, which following \citet{rathborne06} are indicated as G34--MM1 and G34--MM2, respectively. The core G34--MM1, which is the less evolved one, shows conspicuous extended emission at 4.5~$\mu$m and is associated with the position of the EGO G34.41+0.24  \citep[catalogued by][]{cyganowski2008}. By the other hand, the core G34--MM2, which is brighter at 8~$\mu$m, is associated with the UC \hii~region G34.4+0.23. 

Several authors have found molecular outflow activity towards the cores G34-MM1, MM2, and MM3, showing that at least three molecular condensations embedded in IRDC G34 are active star-forming sites \citep{shepherd07, rathborne08, sanhueza10, sakai18}. \citet{shepherd07} found two perpendicular molecular outflows arising from the core G34--MM1, however limitation in the spatial resolution of the data prevented them from the identification of the driving individual sources. Finally, the authors, based on a comparison between the gravitational binding energy and the outflow kinetic energy, suggested that the core G34--MM1 is been destroyed by the intense outflow activity. 

\citet{rathborne08, rathborne2011} pointed out that G34--MM1 is a hot molecular core with a rotating gaseous structure surrounding a central protostar. The authors detected the emission of several COMs lines towards this core, including CH$_{3}$CN (18$_{6}$--17$_{6}$), from which they estimated a rotational temperature between 110 and 210~K.  In addition, based on the ratio between the gas and virial masses of core G34--MM1, \citet{rathborne08} found that the molecular condensation is likely to be fragmented, although no evidence of fragmentation was observed.

More recently, \citet{tang19} carried out a study on the impact of the fragmentation of cores G34--MM1, G34--MM2, and G34--MM3, based on a comparison between the effects of gravity, magnetic field, and turbulence. The authors suggested that the absence of fragmentation in the core G34--MM1 could be due to the high ratio between the gravitational energy density and the turbulent pressure. However, the presence of two perpendicular molecular outflows towards G34--MM1 detected by \citet{shepherd07} would indicate an internal fragmentation of the core that deserves to be studied with higher angular resolution data.

\section{Data}
\label{dataS}

Data cubes with central frequencies at 93 and 334~GHz were obtained from the ALMA Science Archive\footnote{http://almascience.eso.org/aq/} (Project codes: 2015.1.00369, and 2013.1.00960, PIs: Rosero, V., and Csengri, T., respectively) 
to analyze the core G34--MM1. The Common Astronomy Software Applications (CASA) 5.4.1 was used to handle and analyse the data. Table\,\ref{data} presents the main data parameters.  The single pointing observations were carried out using the following telescope configurations: Min/Max Baseline(m) of 150/460 and extended for 93 and 334~GHz, respectively, in the 12~m array in both cases. The maximum recoverable scales at 93 and 334~GHz are 14.5 and 6.7~arcsec, respectively.

Given that high-spatial resolution is required to perform this study, it is important to remark that the beam size of the 334~GHz data cube provides a spatial resolution of about 0.014~pc ($\sim$ 2900 au) at the distance of 3.6~kpc, which is appropriate to investigate the fragmentation of core G34--MM1. 

\begin{figure}[h]
   \centering
   \includegraphics[width=9cm]{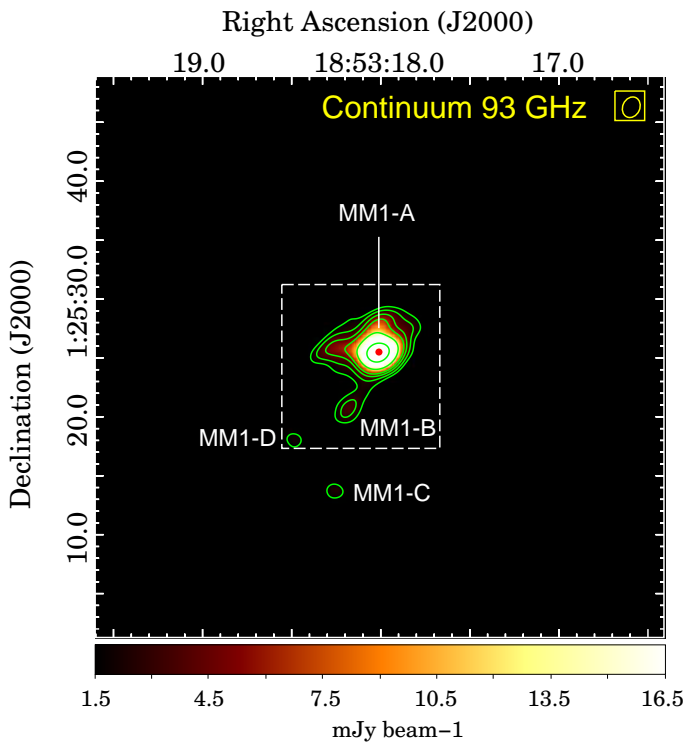}
    \caption{Radio continuum emission at 93~GHz. The colour-scale goes from 1.5 (about 5~$\sigma$ rms noise level) to 16.5~mJy beam$^{-1}$. The contours levels are at 1.5, 2.5, 4, 5, 12, 30~mJy beam$^{-1}$. The red filled circle indicates the position and extension (up to 3~$\sigma$ rms noise level) of the radio continuum emission at 1.3~cm detected by \citet{rosero16}.  The white box indicates the region mapped at 334~GHz.  The  beam  is indicated at the top right corner.}
              \label{93GHz}
    \end{figure}

\begin{table}
\tiny
\centering
\caption{Main data parameters.}
\label{data}
\begin{tabular}{lcccc}
\hline\hline
                  & beam size & FoV & $\Delta$v & rms  Noise          \\
                  &   (\s)             & (arcsec$^2$) & \ks &(mJy beam$^{-1}$)\\
\hline                 
Cont. 93 GHz      &  2.1 $\times$ 1.4  & 45$\times$45 & - & 0.3      \\
HCO$^+$ J=1--0    &  1.9 $\times$ 1.4  & 45$\times$45 & 1.5 &5.0      \\
Cont. 334 GHz     &  0.8 $\times$ 0.7  & 15$\times$15 & - &10       \\
$^{12}$CO J=3--2  &  0.8 $\times$ 0.7  & 15$\times$15 & 0.9 &15       \\

\hline
\end{tabular}
\end{table}

\section{Results}
\label{results}

In the following subsections we present the results obtained based on the analysis of the ALMA data towards the core G34--MM1. 

\subsection{Continuum emission at 93 and at 334~GHz} 

Figure \ref{93GHz} shows the radio continuum emission at 93~GHz.  The lower contour is at 1.5~mJy beam$^{-1}$ (about 5~$\sigma$ rms noise level). The peak of the emission
at 93~GHz (${\rm \lambda} \sim$ 3.2~mm) is positionally coincident with the peaks observed at 1.3 and 6~cm by \citet{rosero16} (red filled circle in the figure).

It can be appreciated a main condensation, named MM1--A,  which exhibits a conspicuous bulge towards the east (increasing RA) and seems to be connected to a smaller condensation, named MM1--B, that extends to the southeast (decreasing DEC.). There are also two minor condensations (MM1--C and MM1--D marginally detected at 5~$\sigma$ rms noise level) which appear detached from the main core and are located towards the southeast from it.

 \begin{figure}[h]
   \centering
   \includegraphics[width=9cm]{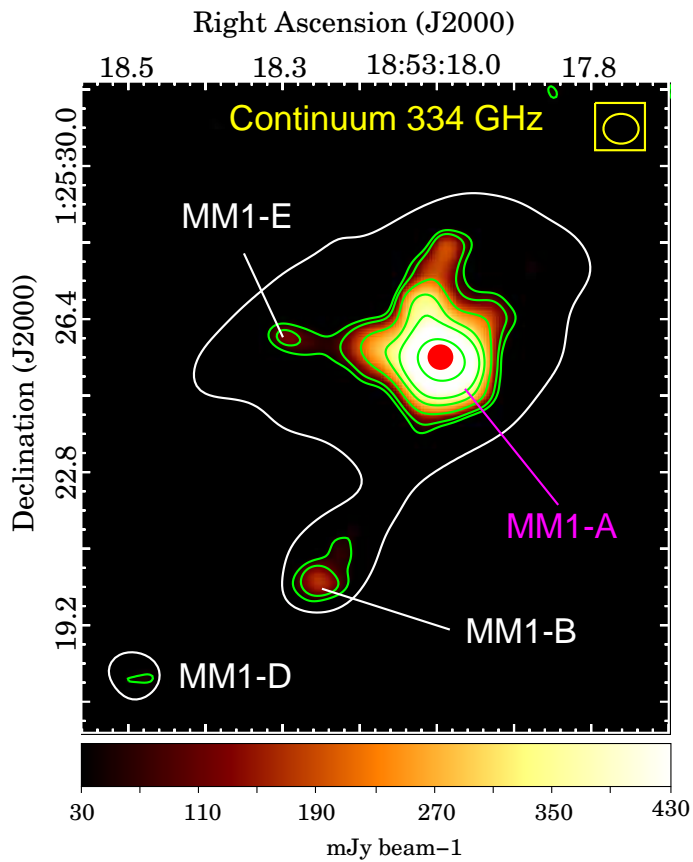}
    \caption{Radio continuum emission at 334~GHz.  The colour-scale goes from 30 (about 3~$\sigma$ rms noise level) to 430~mJy beam$^{-1}$. The green contours levels are at 30, 50 (about 5~$\sigma$ rms noise level), 100, 150, 250, and 700~mJy beam$^{-1}$. The white contour corresponds to the 5~$\sigma$ rms noise level of the radio continuum emission at 93~GHz. The red filled circle indicates the position and extension (up to 3~$\sigma$ rms noise level) of the radio continuum emission at 1.3~cm detected by \citet{rosero16}. The beam is indicated at the top right corner.}
              \label{334GHz}
    \end{figure}


Figure \ref{334GHz} shows the radio continuum emission at 334~GHz (colour and green contours) in comparison with the emission at 93~GHz represented with a  white contour.
The peak of the emission at 3~cm detected by \citet{rosero16} is marked with a red filled circle and coincides with the peak of the emission at 334~GHz ($\lambda \sim$ 0.89~mm).

  \begin{figure*}[h]
   \centering
   \includegraphics[width=18cm]{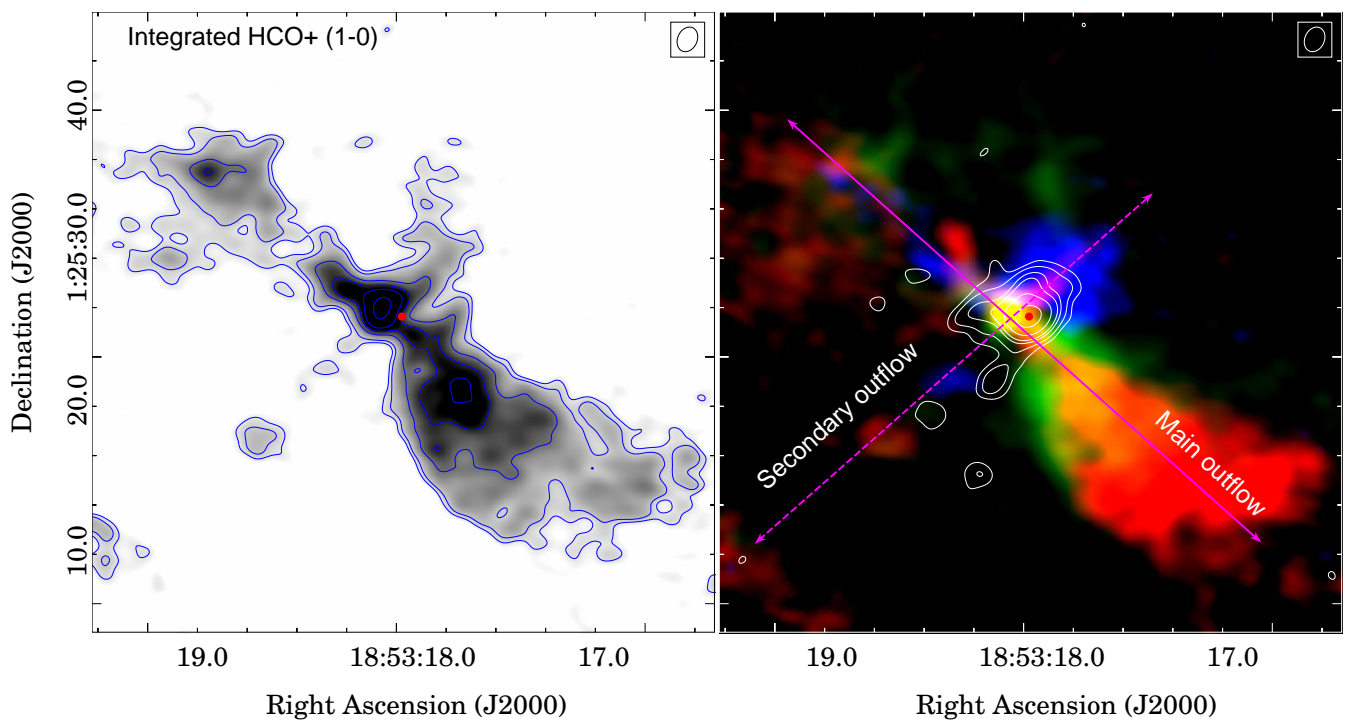}
       \caption{Left panel: HCO$^+$ J=1--0 line emission distribution integrated between 40 and 75~\ks. The gray-scale goes from 0.1 to 0.6 Jy beam $^{-1}$ \ks. The blue contours are at 0.1, 0.2, 0.3, 0.5, 0.6, and 0.9~Jy beam$^{-1}$ \ks.  Right panel: three colour image of the HCO$^+$ J=1--0 line emission integrated over three velocities intervals: [43, 53], [53, 63], and [63, 73]~\ks, shown in blue, green, and red, respectively. The solid and dashed lines indicate the direction of the main outflow and its perpendicular one, respectively. White contours represent the radio continuum emission at 93~GHz. Levels are at 1, 2, 4, 5, 12, 30~mJy beam$^{-1}$. The red filled circle indicates the position of the radio continuum emission at 1.3~cm detected by \citet{rosero16}. The beam is indicated at the top right corner of each panel.}
              \label{hco}
    \end{figure*}

Despite the differences in sensitivity and angular resolution between  images at 93 and 334~GHz, some common features can be noticed.  For instance, the condensation named MM1--B, which at 334~GHz appears detached from the main core coincides with an elongated protuberance at 93~GHz extending southeastwards the core. On the other hand, at 334~GHz, it is noticeable a bridge-like structure extending to the east of the core and ending in a small condensation (hereafter MM1--E) which is unresolved at 93~GHz. In conclusion, the continuum emission at 334 GHz, which essentially traces the core dust emission, spatially resolves structures within the core G34--MM1 that were not resolved in previous works, strongly suggesting that the core is indeed fragmented.


 In what follows we analyse the nature of the continuum emission at both frequencies. \citet{shepherd04} estimated a total flux density of 56.8 and 0.7~mJy at 3~mm and 6~cm, respectively, towards G34--MM1. Considering that continuum emission at 6~cm arises completely from ionized gas and assuming that its emission is optically thin between 6~cm and 3~mm ($S_{\nu}\propto {\nu}^{-0.1}$), the authors estimated a contribution of 0.52~mJy of the ionized gas to the continuum emission at 3~mm (about 1 percent). 

Later, \citet{rosero16} estimated a total flux density of about 0.4 and 1.5~mJy at 4.9~GHz ($\lambda \sim$ 6~cm) and 25.5~GHz ($\lambda \sim$ 1.3~cm), respectively, with a spectral index of $+$0.7, which is in agreement with optically thick thermal emission. Taking into account the location of this ionized gas (red filled circle in Figure \ref{334GHz}), it is reasonable to assume that its contribution to the continuum emission at 93 and 334~GHz will only be relevant towards the center of the condensation MM1--A. Considering a region of the size of a beam (at each frequency) centered at the position of the 1.3~cm emission, we estimated a total flux density of about 20 and 710~mJy at 93 and 334~GHz, respectively. Assuming that the turn over frequency of the radio continuum emission for these young \hii~regions can reach the 100~GHz \citep{yang2021} and considering the spectral index obtained by \citet{rosero16}, we estimated a flux density (related to the emission of the ionized gas) of about 3~mJy for both frequencies. Therefore, the upper limit for the contribution of the emission of the ionized gas to the continuum emission at 93~GHz is about 15 percent (towards the center of MM1--A) and is negligible at 334~GHz.

\subsection{HCO$^+$ J=1--0 line emission}

Figure \ref{hco}-left shows the HCO$^+$ J=1--0 line emission distribution integrated between 40 and 75~\ks. It can be noticed a main HCO$^+$ condensation centered at RA = 18:53:18.0, DEC = +1:25:27.0, on whose border is located the ionized gas (red filled circle) related to the B2-type protostar.
This HCO$^+$ condensation, which seems to be positionally related to MM1--A, appears in connection with several gaseous elongated structures. Taking into account that the HCO$^+$ J=1--0 line is a good tracer of dense gas and molecular outflows (e.g. \citealt{rawlings2004}) and based on the morphology of the emission,  we conclude that what it is displayed in Fig. \ref{hco}-left clearly indicates the presence of two perpendicular outflows arising from MM1--A. These outflows coincide with those previously detected at lower angular resolution ($\sim$4\s) by \citet{shepherd07} using the $^{12}$CO J=1--0 transition. The main molecular outflow, which goes in the southwest-northeast direction, exhibits an asymmetric morphology with respect to its lobes, and reveals a lower degree of collimation than the secondary and perpendicular outflows extending from southeast to northwest.  This secondary outflow also shows a striking asymmetry between its lobes. In particular, the southeastern one exhibits a discontinuous structure composed by at least two fragments.  

   \begin{figure}[h]
   \centering
   \includegraphics[width=9cm]{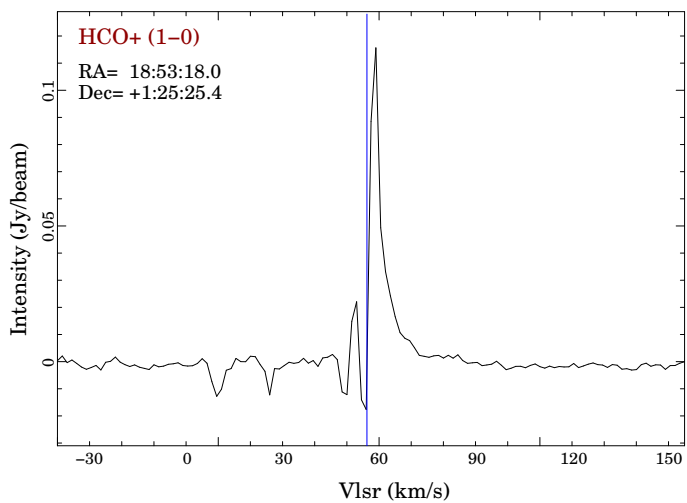}
    \caption{Averaged HCO$^+$ J=1--0 spectrum obtained toward the position of the radio continuum emission at 1.3~cm in a circular region of 2~arcsec of radius. The blue vertical line indicates the systemic velocity of the gas associated with G34--MM1.}
   \label{hco_spectrum}
    \end{figure}

   \begin{figure*}
   \centering
   \includegraphics[width=18cm]{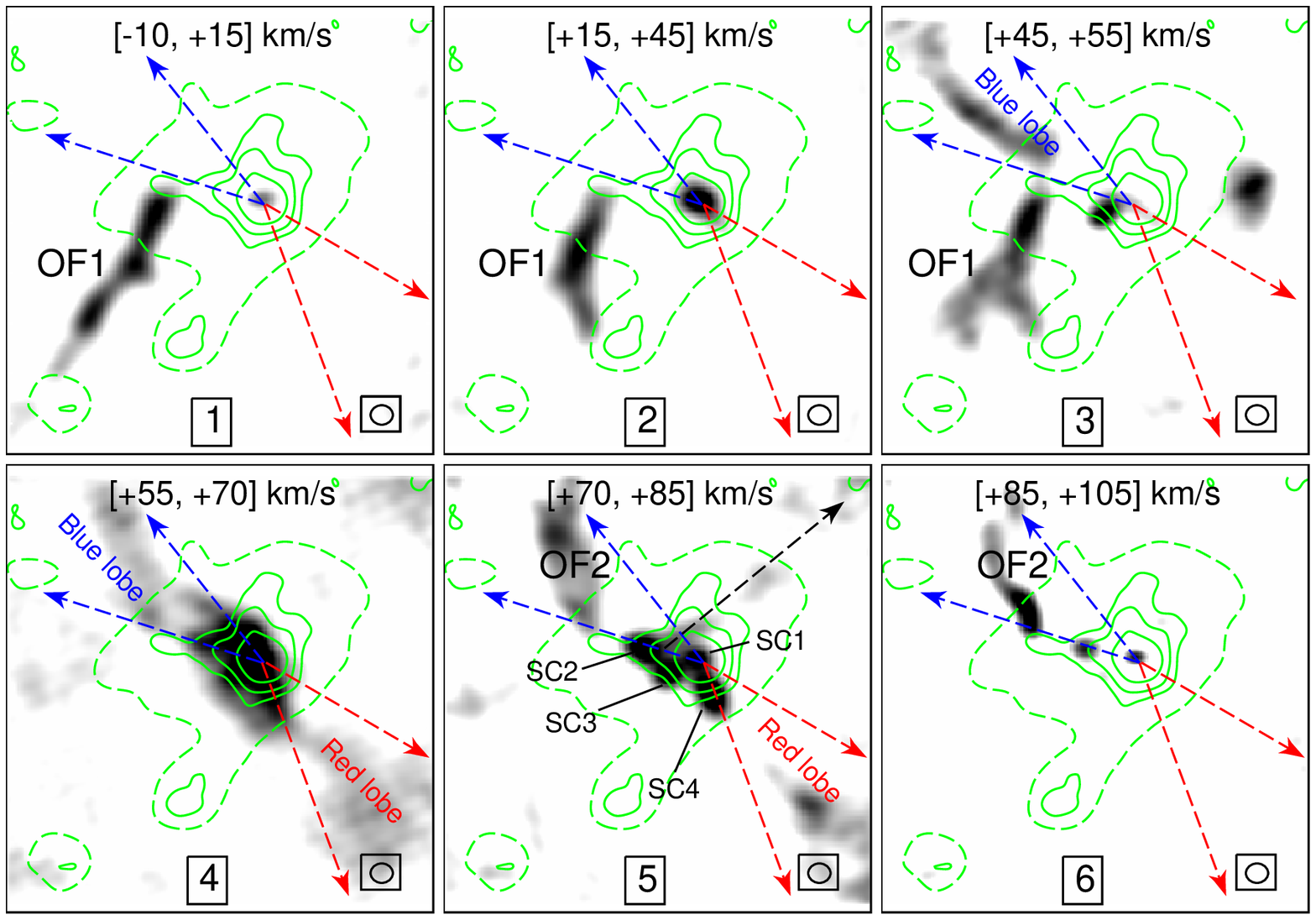}
    \caption{Integrated velocity channel maps of the $^{12}$CO J = 3$-$2 emission. The integration velocity range is exhibited at the top of each panel. Gray-scale goes from 2~Jy beam$^{-1}$~\ks (about 3$\sigma$ rms noise of an averaged channel map) to 10~Jy beam$^{-1}$~\ks. Green contours represent the continuum emission at 334~GHz. Levels are at  30, 200, and 400~mJy beam$^{-1}$.  Dashed contour represents the 3~$\sigma$ rms noise level of the radio continuum emission at 93~GHz. The beam is indicated at the bottom right corner of each map.}
             \label{coint}
    \end{figure*}

After a careful analysis of the HCO$^+$~data cube with the aim of identify different velocity components, we present in Figure \ref{hco}-right, a three-colour image of the HCO$^+$ J=1--0 integrated over three velocity intervals: [43, 53], [53, 63], and [63, 73]~\ks, shown in blue, green, and red, respectively. The central velocity of the second interval corresponds to the systemic velocity of the gas related to G34--MM1 ($\sim$58~\ks). The solid and dashed lines indicate the directions of the main and the secondary molecular outflows, respectively. The main outflow shows a red-shifted lobe with a cone-like morphology extending to the southwest. Interestingly,  this lobe shows the presence of `systemic velocity' gas (seen in green) probably associated with the walls of  a cavity excavated in the surrounding dense gas by the action of the outflow (e.g. \citealt{wei06,arce07}). As mention above, this lobe is noticeably brighter than the northeastern one, which shows the presence of a more collimated, and curved blue-shifted feature, and also some diffuse red-shifted gas almost coexisting in the same spatial region.

The secondary outflow exhibits a southeast-northwest direction. The northwest direction shows two aligned structures: an extended and less collimated blue-shifted emission, which positionally coincides with the blue lobe detected by \citet{shepherd07}; and a more collimated red-shifted filament embedded in the former emission. In turn, towards the southeast two structures appear, a blue- and a red-shifted one, which are aligned with the outflow axis and detached from the molecular core. 

 Figure \ref{hco_spectrum} shows an averaged HCO$^+$ J=1--0 spectrum obtained toward the position of the radio continuum emission at 1.3 cm. Four conspicuous self-absorption features can be appreciated, and the velocity of the main self-absorption dip coincides with the systemic velocity of the gas related to G34--MM1 (blues vertical line in the figure).

\subsection{$^{12}$CO J=3--2 line emission}

Figure \ref{coint} shows in grey-scale the integrated velocity channel maps of the $^{12}$CO J=3$-$2 emission with the radio continuum emission at 93 and 334~GHz superimposed in contours. 

\begin{figure}[h]
   \centering
   \includegraphics[width=9cm]{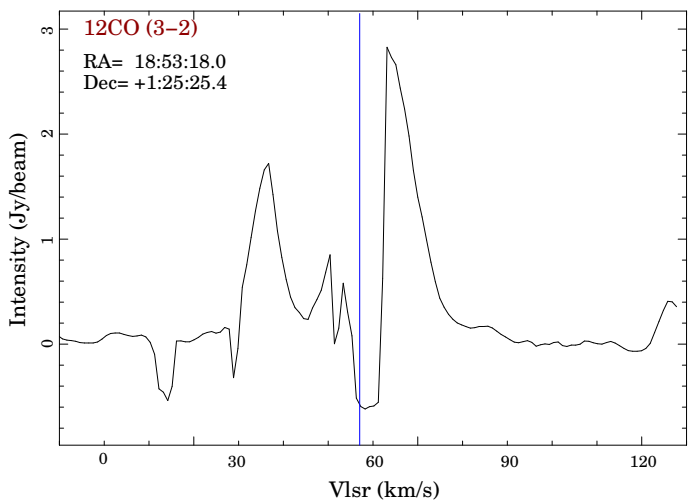}
    \caption{Averaged $^{12}$CO J=3--2 spectrum obtained toward the position of the radio continuum emission at 1.3~cm in a circular region of 1~arcsec of radius. The
blue vertical line indicates the systemic velocity of the gas associated with G34--MM1.}
              \label{co_spectrum}
    \end{figure}

Panel 1 shows a straight filament-like structure (named OF1) of about 0.2~pc in size, which extends from the MM1--E position (see Fig. \ref{334GHz}) to the southeast. In panel 2, the feature OF1 exhibits a curved morphology, which seems to connect condensations MM1--E and MM1--B. It can also be appreciated the molecular gas related to MM1--A. Panel 3 shows the presence of gas very likely related to the blue-shifted lobe associated with the main molecular outflow, which at this velocity range appears disconnected from the main core. The gaseous structure OF1 exhibits a cone-like morphology ending in two separated branches.  The absence of molecular gas towards the location of the radio continuum emission peak at this velocity range is due to self-absorption effects as can be seen in Fig. \ref{co_spectrum}. Panel 4, whose velocity range includes the systemic velocity of G34--MM1, shows the bulk of the $^{12}$CO J=3--2 emission related to MM1--A. The region delimited by the dashed blue lines (blue lobe region) shows the presence of two structures: the low velocity gas component of the blue-shifted lobe belonging to the main outflow, which appears to be connected to the bulk of the emission, and diffuse emission which seems to arise from MM1--E. The red-shifted lobe of the main outflow can also be appreciated in this panel.

In panel 5 there are several interesting features: (1) the high velocity gas component of the red-shifted lobe of the main outflow, (2) a cone-like gaseous structure (indicated as OF2), which seems to arise from the condensation MM1--E (see Fig. \ref{334GHz}), (3) four molecular condensations, likely  subcores, named SC1 through SC4, with SC1 in positional coincidence with the radio continuum emission peak of MM1--A, and (4) a discontinuous filament related to the secondary outflow whose direction is indicated with the dashed black line, which seems to arise from the intermediate region between subcores SC2 and SC3. Finally, panel 6 shows the gaseous structure OF2, which appears more collimated than in panel 5. Given their morphology and kinematics, both gaseous structures OF1 and OF2, appear to be high velocity molecular outflows related to the dust condensation MM1--E. It is important to mention that these new molecular outflows were not identified in the HCO$^+$ J=1--0 emission.

 Figure \ref{co_spectrum} shows the averaged $^{12}$CO J=3--2 spectrum obtained toward the position of the radio continuum emission at 1.3~cm. As in the case of the HCO$^+$ J=1--0 spectrum, it can be appreciated several self-absorption features. In particular, the velocity of the main self-absorption dip coincides with the systemic velocity of the gas related to G34--MM1. It is important to mention that these pronounced self-absorption features are only detected towards the center of the MM1--A condensation.

\subsubsection{High velocity molecular outflows related to MM1--E}
\label{outpar}

\begin{figure}[h]
   \centering
    \includegraphics[width=9cm]{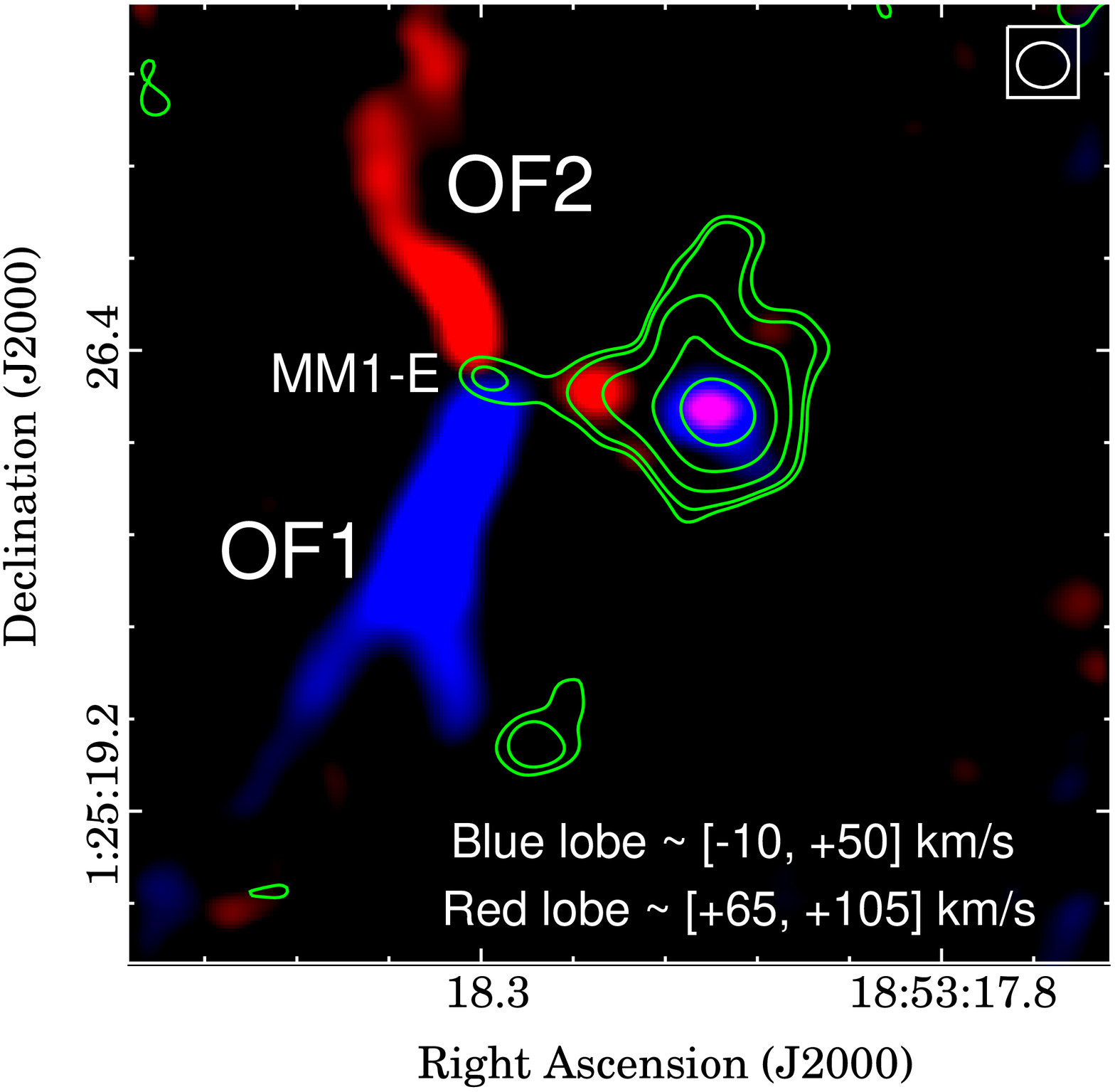}
    \caption{$^{12}$CO J=3--2 emission distribution integrated between $-$10 and +50~\ks~ (feature OF1 seen in blue) and between +65 and +105~\ks~ (feature OF2 seen in red). Green contours represent the radio continuum emission at 334 GHz. Levels are at 30, 50, 100, 250, 700~mJy beam$^{-1}$.} The beam is indicated at the top right corner.
              \label{Nout}
    \end{figure}

Figure \ref{Nout} displays the $^{12}$CO J=3--2 emission distribution integrated over two velocity ranges with the continuum emission at 334~GHz (green contours) superimposed. It can be appreciated two conspicuous filament-like features: OF1 in blue and OF2 in red, associated with the velocity ranges $-$10 to $+$50~\ks, and $+$65 to $+$105~\ks~, respectively. Both gaseous structures seem to arise from  MM1--E. We suggest that they correspond to the misaligned blue- and red-shifted lobes of a high velocity molecular outflow. Therefore, MM1--E would be another active star formation condensation within G34--MM1.  

In order to roughly estimate the mass of the outflow candidates arising from MM1--E, following \citet{bertsch93}, we calculate the H$_2$ column density from:

\begin{equation}
{\rm N(H_2)=2.0 \times 10^{20} \frac{{\rm W(^{12}CO)}}{K~kms^{-1}}},
\label{nh2}
\end{equation}

\noindent where W($^{12}$CO) is the $^{12}$CO J=3--2 integrated intensity at the the corresponding velocity intervals. The W($^{12}$CO) units were converted using:

\begin{equation}
{\rm T[K]=1.22 \times 10^3  \frac{I[mJy/beam]}{\nu^2[GHz] ~\theta_{maj}[arcsec]~\theta_{min}[arcsec]}}.   
\label{mJytoK}
\end{equation}

\noindent where $\theta_{maj}$ and $\theta_{min}$ correspond to the major and minor axis of the beam, respectively. Then, the mass of each outflow was derived from:

\begin{equation}
{\rm M=\mu~m_H~D^2~\Omega\sum_{i}~N_i(H_2)},
\label{nh2b}
\end{equation}

\noindent where $\Omega$ is the solid angle subtended by the beam size, ${\rm m_H}$ is the hydrogen mass, ${\rm \mu}$ is the mean molecular weight, assumed to be 2.8 by taking into account a relative helium abundance of 25 \%, and D is the distance. Summation was performed over all beam positions belonging to the lobes shown in Fig. \ref{Nout}.  Table \ref{outflow} shows the length, the mass, the momentum (${\rm P = M \bar v}$), the energy (${\rm E = 1/2 M \bar v^2}$), the dynamical age (${\rm t_{dyn} = Length/v_{max}}$), and the outflow rate (${\rm \dot M = M /t_{dyn}}$) of each lobe, where ${\rm \bar v}$ and ${\rm v_{max}}$ are the median and maximum velocities of each interval velocity with respect to the systemic velocity of the gas associated with G34-MM1, respectively. 

\begin{table}
\centering
\caption{Main parameters of the molecular outflow related to MM1--E.}
\label{outflow}
\begin{tabular}{ccc}
\hline\hline
                  & Red lobe & Blue lobe \\
\hline\hline                 
Length (pc)       & 0.090 &  0.130  \\
Mass (${\rm M_{\odot}}$)  & 0.8 & 4.1  \\
Momentum (${\rm M_{\odot}~km~s^{-1}}$) & 22 & 156 \\
Energy (10$^{46}$ erg) & 0.5 & 5.6 \\
Dynamical age (10$^3$ yrs) & 1.6 & 1.7 \\
Outflow rate ($10^{-3}~{\rm M_{\odot}~yr^{-1}}$) & 0.5 & 2 \\

\hline
\end{tabular}
\end{table}

Finally, taking into account the non-detection of molecular gas related to MM1--E in the $^{12}$CO J=3--2 emission, and considering that this dust condensation is marginally detected at 334~GHz, we can not discard the possibility that gaseous structures OF1 and OF2, indeed correspond to filaments of molecular gas converging to MM1--E from the unbound gas of the clump, as several works have suggested towards other cores \citep[e.g.,][]{sch19}.

\section{Discussion}
\label{discuss}

 In this section we discuss the main results of our analysis regarding evidences of fragmentation within the molecular core G34--MM1, and the star formation activity in the region.

Previous studies  \citep[e.g.,][]{shepherd04, rathborne08, sanhueza10} have not found any sign of fragmentation associated with G34--MM1. In this work, using the high sensitivity and angular resolution of the ALMA data,  we suggest that the fragmentation process could have occurred within this molecular core, at least, at two different spatial scales. The larger spatial scale fragmentation within the core G34--MM1 can be appreciated in the radio continuum image at 93~GHz (see Fig.\,\ref{93GHz}), which shows evidence of fragmentation at core scale with a main condensation MM1--A and three minor fragments (MM1--B, MM1--C, and MM1--D, with the last two marginally resolved). The average spatial separation among the main condensation and these fragments is about 0.2~pc. This main condensation seen at 93~GHz is resolved into two dust condensations at 334~GHz, MM1--A and MM1--E (see Fig.\,\ref{334GHz}). 

 In Sect.\,\ref{outpar} the main parameters of the molecular outflow associated with MM1--E were estimated. A comparison of our results with those presented in recent works carried out with ALMA \citep[for example;][]{cyg2017, hervias2019, li2020}, suggests that the source embedded in MM1--E is generating a young (${\rm t_{dyn}} \sim 1.6 \times 10^3$~yrs), massive (M $\sim$ 5 M$_{\odot}$) and energetic  (E $\sim 6 \times 10^{46}$~ergs) molecular outflow.  

\begin{figure}[h]
   \centering
    \includegraphics[width=9cm]{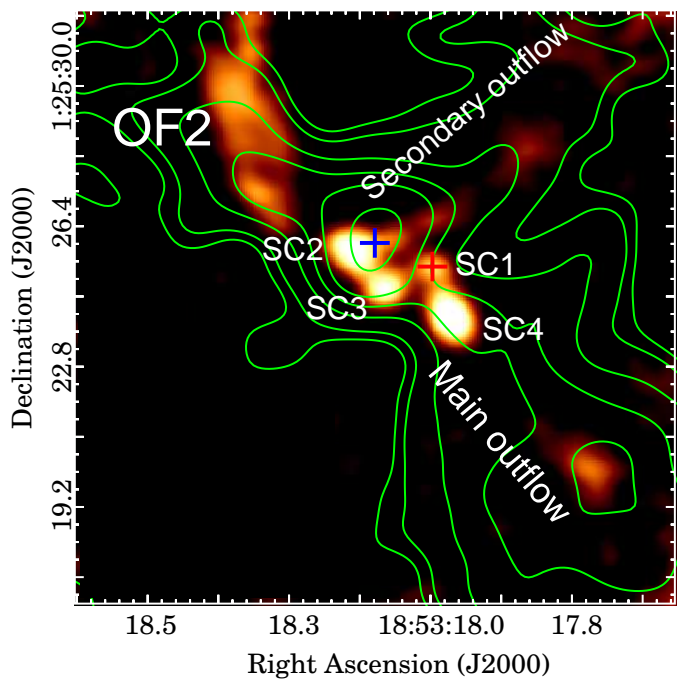}
        \caption{$^{12}$CO J=3--2 emission distribution integrated between +70 and +75~\ks. This  velocity range was chosen to highlight the secondary outflow arising from SC2. Contours correspond to the HCO$^+$ J=1--0 emission distribution integrated between 40 and 75~\ks. Levels are at 0.1, 0.2, 0.3, 0.5, 0.6, and 0.9~Jy beam$^{-1}$ \ks. The red cross indicates the peak of the radio continuum emission at 1.3~cm \citep{rosero16}.  The blue cross represents the position of the emission peak of the integrated HCO$^+$ J=1--0 emission.}
              \label{SChco+}
    \end{figure}

 From the analysis of the $^{12}$CO J=3--2 emission, we also found possible evidence of internal fragmentation within MM1--A (see mainly panel 5 in Fig.\,\ref{coint}). To better appreciate such fragmentation, in Fig.\,\ref{SChco+} we show the $^{12}$CO J=3--2 emission distribution integrated between 70 and 75~km s$^{-1}$, the velocity range in which the presence of subcores candidates (SC1 to SC4) is more evident. The average size of these subcores is about 0.02~pc and the average spatial separation among them is about 0.03~pc.

The subcore SC1 positionally coincides with the peak of the emission at 1.3~cm (red cross, from \citealt{rosero16}). Hence, we suggest that in SC1 may be embedded the driving source of the main molecular outflow observed at both, HCO$^+$ and $^{12}$CO emissions.

Regarding the other three molecular condensations we found that the position of the peaks of SC2 and SC3 does not shift, in more than a beam, throughout their velocity range, which suggest that both subcores are indeed evidence of the internal fragmentation of MM1--A. In particular, SC2 is the closest subcore to the main HCO$^+$ central feature (blue cross in Fig.\,\ref{SChco+}). Additionally, it seems to be connected with a conspicuous filament-like structure extending towards the northwest, in spatially coincidence with the lobe of the secondary outflow traced by the HCO$^+$ emission. Thus, we suggest that in the subcore SC2 may be embedded the source of the secondary molecular outflow.  

Finally, given that there is no evidence of molecular outflow activity towards the subcore SC3, we can not discard that this condensation corresponds to shocked gas associated with the secondary molecular outflow.
In the case of SC4, Fig.\,\ref{SChco+} shows that it has a slightly elongated morphology aligned with the symmetry axis of the red lobe of the main molecular outflow, and it spatially coincides with the region in which the outflow lobe originates. The position of the emission peak of SC4 shifts about 0.8~arcsec (the beam size) along its symmetry axis, throughout the velocity range, suggesting that it would be associated with shocked gas in the region where the outflow arises.

The evidence of fragmentation within a hot molecular core is something scarce in the literature. It can be mentioned, for instance, the work of \citet{dall19}, who observed and analyzed many cores and subcores in the star-forming complex G9.62$+$0.19.   
In our work, the hot core G34--MM1 shows strong evidence of fragmentation with some of the molecular condensations exhibiting molecular outflow activity (SC1, SC2, and MM1--E), suggesting the presence of several protostars embedded within it.    

Our results, as it was suggested by \citet{die2015} for the region of Mon R2, would support the competitive accretion scenario of star formation, which can explain the distribution of stellar masses, the mass segregation of young stellar clusters, and the high binary frequency and properties of massive stars \citep{bonnell08}.

\section{Concluding remarks} 
\label{concl}

We carried out a study of the internal structure of the hot molecular core G34--MM1 embedded in the IRDC G34.43+00.24, using high resolution ALMA data centered at 93 and 334~GHz.

 We found evidence of core fragmentation at two different spacial scales. In particular, the internal fragmentation of the condensation MM1--A, allows us to identify the molecular subcores candidates to harbour the protostars responsible for the generation of two perpendicular molecular outflows.  

 The marginally detected dust condensation MM1--E appears to be connected with two filament-like gaseous structures, which would be tracing a young, massive and energetic molecular outflow. 

 The fragmentation of the hot molecular core G34--MM1 at two different spatial scales, together with the presence of multiple molecular outflow associated with it, would support the competitive accretion hypothesis of star formation.

\begin{acknowledgements}

We acknowledge the anonymous referee for her/his helpful comments  and  suggestions.
N.I. and M.B.A. are posdoctoral and doctoral fellows of CONICET, Argentina, respectively. 
S.P. and  M.O. are members of the Carrera del Investigador Cient\'\i fico of CONICET, Argentina. 

This work is based on the following ALMA data: ADS/JAO.ALMA $\#$ 2015.1.00369, and 2013.1.00960. ALMA is a partnership of ESO (representing its member states), NSF (USA) and NINS (Japan), together with NRC (Canada), MOST and ASIAA (Taiwan), and KASI (Republic of Korea), in cooperation with the Republic of Chile. The Joint ALMA Observatory is operated by ESO, AUI/NRAO and NAOJ.

\end{acknowledgements}

%
%

\bibliographystyle{aa}  
\bibliography{ref}
\IfFileExists{\jobname.bbl}{}
{\typeout{}
\typeout{****************************************************}
\typeout{****************************************************}
\typeout{** Please run "bibtex \jobname" to optain}
\typeout{** the bibliography and then re-run LaTeX}
\typeout{** twice to fix the references!}
\typeout{****************************************************}
\typeout{****************************************************}
\typeout{}
}
\label{lastpage}

\end{document}